\input harvmac
\def\ev#1{\langle#1\rangle}
\input amssym
\input epsf
\def\unit{\relax{\rm 1\kern-.26em I}}
\def\nada{\relax{\rm 0\kern-.30em l}}
\def\slashed#1{\setbox0=\hbox{$#1$}           
   \dimen0=\wd0                                 
   \setbox1=\hbox{/} \dimen1=\wd1               
   \ifdim\dimen0>\dimen1                        
      \rlap{\hbox to \dimen0{\hfil/\hfil}}      
      #1                                        
   \else                                        
      \rlap{\hbox to \dimen1{\hfil$#1$\hfil}}   
      /                                         
   \fi}



\noblackbox
\def\IL{\relax{\rm I\kern-.18em L}}
\def\IH{\relax{\rm I\kern-.18em H}}
\def\IR{\relax{\rm I\kern-.18em R}}
\def\IC{\relax\hbox{$\inbar\kern-.3em{\rm C}$}}
\def\IZ{\relax\ifmmode\mathchoice
{\hbox{\cmss Z\kern-.4em Z}}{\hbox{\cmss Z\kern-.4em Z}}
{\lower.9pt\hbox{\cmsss Z\kern-.4em Z}} {\lower1.2pt\hbox{\cmsss
Z\kern-.4em Z}}\else{\cmss Z\kern-.4em Z}\fi}

\def\CL {{\cal L}}

\def\CO {{\cal O}}

\def\CH {{\cal H}}

\def\CA{{\cal A}}


\def\CO {{\cal O}}

\def\CQ {{\cal Q }}

\def\Tr{{\rm Tr}}

\font\manual=manfnt \def\dbend{\lower3.5pt\hbox{\manual\char127}}

\def\IZ{\relax\ifmmode\mathchoice
{\hbox{\cmss Z\kern-.4em Z}}{\hbox{\cmss Z\kern-.4em Z}}
{\lower.9pt\hbox{\cmsss Z\kern-.4em Z}} {\lower1.2pt\hbox{\cmsss
Z\kern-.4em Z}}\else{\cmss Z\kern-.4em Z}\fi}
\def\half {{1\over 2}}

\def\lfm#1{\medskip\noindent\item{#1}}

\def\bar{\overline}

\def\CH{{\cal H}}

\def\RHS{right hand side}

\def\frac#1#2{{{#1}\over{#2}}}
\lref\PolchinskiDY{
  J.~Polchinski,
  ``SCALE AND CONFORMAL INVARIANCE IN QUANTUM FIELD THEORY,''
  Nucl.\ Phys.\  B {\bf 303}, 226 (1988).
}
\lref\GW{D.~J.~Gross and F.~Wilczek,
``Asymptotically Free Gauge Theories. 2,''
Phys.\ Rev.\ D {\bf 9}, 980 (1974).
}
\lref\HoldomSK{
  B.~Holdom,
  ``Techniodor,''
  Phys.\ Lett.\  B {\bf 150}, 301 (1985).
}
\lref\BZ{T.~Banks and A.~Zaks,
``On The Phase Structure Of Vector - Like Gauge Theories With Massless Fermions,''
Nucl.\ Phys.\ B {\bf 196}, 189 (1982).
}
\lref\ColemanAD{
  S.~R.~Coleman and J.~Mandula,
  ``ALL POSSIBLE SYMMETRIES OF THE S MATRIX,''
  Phys.\ Rev.\  {\bf 159}, 1251 (1967).
}
\lref\Mack{
  G.~Mack,
  ``All Unitary Ray Representations Of The Conformal Group SU(2,2) With
  Positive Energy,''
  Commun.\ Math.\ Phys.\  {\bf 55}, 1 (1977).
}
\lref\GeorgiEK{
  H.~Georgi,
  ``Unparticle Physics,''
  Phys.\ Rev.\ Lett.\  {\bf 98}, 221601 (2007)
  [arXiv:hep-ph/0703260].
}
\lref\StrasslerBV{
  M.~J.~Strassler,
  ``Why Unparticle Models with Mass Gaps are Examples of Hidden Valleys,''
  arXiv:0801.0629 [hep-ph].
}
\lref\GeorgiSI{
  H.~Georgi,
  ``Another Odd Thing About Unparticle Physics,''
  Phys.\ Lett.\  B {\bf 650}, 275 (2007)
  [arXiv:0704.2457 [hep-ph]].
}
\lref\CheungZZA{
  K.~Cheung, W.~Y.~Keung and T.~C.~Yuan,
  ``Collider signals of unparticle physics,''
  Phys.\ Rev.\ Lett.\  {\bf 99}, 051803 (2007)
  [arXiv:0704.2588 [hep-ph]].
}
\lref\MinwallaKA{
  S.~Minwalla,
  ``Restrictions imposed by superconformal invariance on quantum field
  theories,''
  Adv.\ Theor.\ Math.\ Phys.\  {\bf 2}, 781 (1998)
  [arXiv:hep-th/9712074].
}
\lref\FoxSY{
  P.~J.~Fox, A.~Rajaraman and Y.~Shirman,
  ``Bounds on Unparticles from the Higgs Sector,''
  Phys.\ Rev.\  D {\bf 76}, 075004 (2007)
  [arXiv:0705.3092 [hep-ph]].
}
\lref\BanderND{
  M.~Bander, J.~L.~Feng, A.~Rajaraman and Y.~Shirman,
  ``Unparticles: Scales and High Energy Probes,''
  Phys.\ Rev.\  D {\bf 76}, 115002 (2007)
  [arXiv:0706.2677 [hep-ph]].
}
\lref\NakayamaQU{
  Y.~Nakayama,
  ``SUSY Unparticle and Conformal Sequestering,''
  Phys.\ Rev.\  D {\bf 76}, 105009 (2007)
  [arXiv:0707.2451 [hep-ph]].
}

\lref\AlcarazMX{
  J.~Alcaraz {\it et al.}  [ALEPH Collaboration],
  ``A combination of preliminary electroweak measurements and constraints on
  the standard model,''
  arXiv:hep-ex/0612034.
}
\lref\GoldbergTT{
  H.~Goldberg and P.~Nath,
  ``Ungravity and Its Possible Test,''
  arXiv:0706.3898 [hep-ph].
}
\lref\RivaGD{
  V.~Riva and J.~L.~Cardy,
  ``Scale and conformal invariance in field theory: A physical
  counterexample,''
  Phys.\ Lett.\  B {\bf 622}, 339 (2005)
  [arXiv:hep-th/0504197].
}
\lref\OsbornCR{
  H.~Osborn and A.~C.~Petkou,
  ``Implications of conformal invariance in field theories for general
  dimensions,''
  Annals Phys.\  {\bf 231}, 311 (1994)
  [arXiv:hep-th/9307010].
}
\lref\CohenSQ{
  A.~G.~Cohen and H.~Georgi,
  ``Walking Beyond The Rainbow,''
  Nucl.\ Phys.\  B {\bf 314}, 7 (1989).
}
\lref\AppelquistDQ{
  T.~Appelquist, J.~Terning and L.~C.~R.~Wijewardhana,
  ``The Zero Temperature Chiral Phase Transition in SU(N) Gauge Theories,''
  Phys.\ Rev.\ Lett.\  {\bf 77}, 1214 (1996)
  [arXiv:hep-ph/9602385].
}
\newbox\tmpbox\setbox\tmpbox\hbox{\abstractfont }
\Title{\vbox{\baselineskip12pt \hbox{UCSD-PTH-08-01}}}
{\vbox{\centerline{Comments on Unparticles}}}
\smallskip
\centerline{Benjamin Grinstein$^{1}$, Kenneth Intriligator$^{1}$ and Ira Z.  Rothstein$^2$}
\smallskip
\bigskip
\centerline{$^1${\it Department of Physics, University of
California, San Diego, La Jolla, CA 92093 USA}}
\medskip
\centerline{$^2${\it Physics Department, Carnegie Mellon University, Pittsburgh PA 15213, USA}}
\bigskip
\vskip 1cm

\noindent 
We comment on several points concerning unparticles
which have been overlooked in the literature.  One regards Mack's
unitarity constraint lower bounds on CFT operator dimensions, {\it e.g.,} $d_V\geq 3$ for
primary, gauge invariant, vector unparticle operators.  We correct the
results in the literature to account for this, and also for a needed
correction in the form of the propagator for vector and tensor
unparticles.  We show that the unitarity constraints can be directly
related to unitarity requirements on scattering amplitudes of
particles, {\it e.g.,}  those of the standard model, coupled to the CFT
operators.  We also stress the existence of explicit standard model contact
terms, which are generically  induced by the coupling to the CFT (or any
other hidden sector), and are subject to LEP bounds.  Barring an
unknown mechanism to tune away these contact interactions, they can
swamp interference effects generated by the CFT.  We illustrate
these points in the context of a weakly coupled CFT example.  A significant
amount of the unparticle literature should be reconsidered or revised in light of the 
observations in this note. 
\bigskip

\Date{January 2008}

\newsec{Introduction} 

Observed, known particle physics is based on theories which have a
mass gap and/or are free in the infrared.  On the other hand, certain
other theories -- interacting conformal field theories -- behave
differently in the infra-red.  Such theories have a traceless
stress-energy tensor\foot{In principle, a theory could be scale, but
not conformal invariant, if $T_\mu ^\mu$ is a total divergence, rather
than zero.  In practice, however, scale invariant theories are quite generally
also conformal \PolchinskiDY\ -- the only known counterexamples are in $D=2$ spacetime
dimensions, and have other pecularities: non-unitarity \RivaGD,   or non-existence of operator 2-point functions \PolchinskiDY. In particular, the Banks-Zaks \BZ\ type theories have, beyond
scale invariance, symmetry under the full conformal group \PolchinskiDY.}, with all coupling constants at fixed point values,
$g_i=g_i^*$, where the beta functions vanish.  Such theories do not
have a traditional S-matrix description\foot{This is how the conformal 
extension of the Poincare group evades the theorem of \ColemanAD.},
because they do not have free, asymptotically separated in and out
states.  The possibility that certain gauge theories could have an interacting, non-Abelian Coulomb phase has a long history \GW, {\it
e.g.,} at weak coupling in theories which are barely asymptotically
free \BZ.  In the context of supersymmetric theories,
interacting conformal theories are also known to be quite common, and
not limited to weak coupling.  Over the years, there have been many
proposed applications of renormalization group flows which approach  near an interacting CFT in beyond the
Standard Model model building, {\it e.g.,} walking technicolor.

A general class of extensions of the Standard Model involve coupling
the visible sector to an otherwise hidden sector by some ultra-heavy
fields of mass $M$.  Integrating out the ultra-heavies induces higher
dimension operators of the form 
\eqn\vishid{\CL \supset {c_{vh}\over
M^{d_v+d_h-4}}\CO _v\CO _h+{c_{vv}\over M^{2d_v-4}}\CO _v\CO
_v+{c_{vv,1}\over M^{2d_v-2}}\CO _v\partial ^2\CO _v+\dots,} 
where $d_v$ is the operator dimension of the visible sector
operator, $d_h$ is that of the hidden sector operator, and $c_{vh}$
and the other coefficients are dimensionless couplings. In recent
work, Georgi \GeorgiEK\ considered the possibility of such a coupling
to an interacting CFT, 
\eqn\smcft{\CL \supset {\lambda \over M^{d_1+d_2-4}}\CO _{SM} \CO
_{CFT},} 
where $\lambda$ is dimensionless, $d_1=\Delta (\CO _{SM})$ is
the scaling dimension of the operator containing Standard Model fields
and $d_2=\Delta (\CO _{CFT})$ is the dimension of the CFT operator,
including its anomalous dimension, at the CFT renormalization group
fixed point.  The novel aspect is the possibility of unusual $d_2$
values, and its effect on Standard Model scattering amplitudes.
Some possible couplings, with both operators in \smcft\ Lorentz scalars, vectors,
and tensors, were discussed in \GeorgiEK.  Subsequent works explored other
novel aspects of couplings to Lorentz vectors (especially the non-primary case $\CO _{CFT}^\mu =\partial ^\mu \CO _{CFT})$  \refs{\GeorgiSI, \CheungZZA}, scalar \FoxSY, spinor, and tensor CFT operators \GoldbergTT, and other interesting possible signatures \refs{\BanderND, \StrasslerBV}. 

The operators of a CFT are subject to general lower  bounds on their
scaling dimensions, from unitarity \Mack: gauge invariant primary
operators have scaling dimension 
\eqn\dineq{d\geq j_1+j_2+2-\delta _{j_1j_2,0},} 
where $(j_1,j_2)$ are the operator Lorentz spins.    In particular,
gauge invariant\foot{We stress that \dineq\ applies only for gauge invariant operators.
For example, a vector potential $A_\mu$ has $d=1$, which does not contradict \dineq\ because
$A_\mu$ is not gauge invariant.}  primary\foot{Primary means not a derivative of another
operator.  Since scalar operators have $d_S\geq 1$ \dineq, the
non-primary vector operator $\partial _\mu \CO $ has $d\geq 2$, as
opposed to $d_V\geq 3$ for primary vectors.}  vector operators
$\CO^\mu$ have $d_V\geq 3$, with $d_V=3$ if and only if the operator is
a conserved current, $\partial _\mu \CO ^\mu =0$.  We review the
constraints \dineq, and note that much of the unparticle literature has focused on $d$'s in a
range which violate \dineq.  (The latter remark also appears in  \NakayamaQU.)
The unitarity constraints were originally
derived via a quite formal analysis.   Below we   provide a more physical
description of how violations of unitarity arise  in  Standard Model scattering  amplitudes, if they are coupled, as in 
\smcft,  to unitarity violating operators.

We reconsider vector unparticles, with $d$ in the range allowed by
\dineq\ and correct the form of the unparticle propagator
for vector (and tensor) operators, to account for the fact that  
$\partial _\mu \CO ^\mu \neq 0$ if $d\neq 3$.  We also discuss the
situation for integer scaling dimension $d$, which leads to $\log$s in
momentum space and the necessity of a local counter-term.

As indicated in \vishid, coupling the Standard Model to another sector
generally also induces Standard Model contact interactions $c_{vv}\neq
0$, which are subject to experimental bounds \AlcarazMX.  Most of the
unparticle literature omits such contact interactions.  Experimental
constraints from effective contact interactions were first discussed
in \BanderND.  We here note that, in addition to the effective contact
interaction associated with the $c_{vh}$ term in \vishid, explicit
contact interaction terms (the local operators associated with the
$c_{vv}$ terms in \vishid) are generically also generated in the
effective Lagrangian.  Including these explicit contact terms can
easily swamp interference effects arising from CFT interactions.  Even
if some unknown mechanism or fine tuning eliminates the explicit
contact interactions at the scale $M$, they will generally still be
generated at lower energy scales.

\newsec{Operator 2-point functions in CFT and unitarity}

As mentioned in footnote 1, scale invariant theories are generally also conformally invariant. 
In this section, we discuss unitarity constraints on 4d CFTs, and in
particular the result \dineq, which was originally obtained in the
work \Mack.
Representations start with a {\it primary} operator, that is to say an
operator which can not be written as $P_\mu = i\partial _\mu$ on
another operator.  The other operators in the representation are {\it
descendants} of the primary operator, obtained by acting on the
primary operator with $P_\mu$. It is useful to focus on the primary
operators, as results for descendants generally follow from acting
with $P_\mu = i\partial _\mu$.

Conformal invariance completely determines the operator 2-point
functions, in terms of their operator scaling dimensions, up to
operator normalization constants.  We quote expressions for primary
operators.  Two point functions of primary operators are only
non-vanishing if the two operators have the same scaling dimension $d$
(and of course also the same spins $(j_1, j_2)$).  The two-point
function of primary Lorentz scalar operators $\CO$ of dimension
$d=\Delta (\CO )$ with their hermitian conjugates is
\eqn\scalariipt{\ev{\CO (x)^\dagger \CO (0)}=C_S\, {1\over(2\pi)^2}\,{1\over (x^2)^d},} 
where $C_S$ is a constant.  For primary vector operators $\CO _\mu$, of dimension
$d$, the 2-point functions are again determined, up to a normalization constant $C_V$ \OsbornCR\ 
\eqn\vectiipt{\ev{\CO _\mu
(x)^\dagger \CO _\nu (0)}=C_V\, {1\over(2\pi)^2}\,{I_{\mu \nu}(x)\over (x^2)^d},\qquad I_{\mu
\nu}\equiv g_{\mu \nu}-2{x_\mu x_\nu\over x^2}.}  
The particular form of $I_{\mu \nu}$
is completely determined by the conformal symmetry, in particular the
special conformal transformation.\foot{It suffices to consider an
infinitesimal special conformal transformation $x^\mu \to x'{}^\mu
=x^\mu (1+2a \cdot x)-a^\mu x^2$.  A scalar operator of dimension $d $
transforms as $\CO (x)\to \CO '(x)=(1+2a\cdot x)^d \CO (x')$, and a
primary vector operator as $\CO _\mu \to \CO '_\mu = (1+2a\cdot
x)^d(g_{\mu \nu}+2a^\mu x^\nu -2a^\nu x^\mu)\CO _\nu (x')$.  (Vector
descendant operators, $\CO _\mu =\partial _\mu \CO$ transform
differently.)  Writing $\ev{T\CO _\mu (x)\CO _\nu (0)}=(Ag_{\mu
\nu}+Bx_\mu x_\nu/x^2)/(x^2)^d$, invariance under the transformation
requires $B=-2A$.}  For primary symmetric traceless, or antisymmetric,
2-index tensor operators of dimension $d$, the two-point functions are
\OsbornCR, 
\eqn\tensoriipt{\ev{\CO _{\mu \nu}(x)^\dagger \CO _{\lambda
\sigma}(0)}=C_T\, {1\over(2\pi)^2}\,{{\left(I_{\mu \lambda}(x)I_{\nu \sigma}(x)-{1\over 4}
g_{\mu \nu}g_{\lambda \sigma}\right)\pm \mu\leftrightarrow\nu
}\over{ (x^2)^d}}.}  
Similar expressions can be written for higher
tensor representations, and spinor representations can be written in
terms of Dirac $\gamma ^\mu$ matrices.

Unitarity requires that the constants appearing in the above be
positive, \eqn\cpos{C>0,} and the operators can then be rescaled to
set the $C=1$ if we like.  The usual CFT argument for \cpos\ follows
from working with a Euclidean version of the theory, on $S^3\times
\IR$, employing radial quantization, and mapping the operators to
states via $\CO (x\to 0) |0\rangle \to |\CO\rangle$.  Bras are obtained from the kets
in this formalism by acting with inversions, so $\langle 0|\CO
(\infty) ^\dagger \to \langle \CO |$. It then follows that
$C\propto || |\CO \rangle ||>0$.  The unitarity conditions \cpos\ lead
to the unitarity conditions of \Mack\ on the operator dimensions $d$.
The idea is that even if a primary satisfies the positivity conditions
\cpos, a descendant can violate it.  This leads to the operator
dimension requirement \dineq\ of \Mack, to avoid having negative norm
first descendants (acting with a single $P_\mu$) for non-scalar
operators, or second descendants (acting with $P_\mu P^\mu$) for
scalar operators.  See \MinwallaKA\ for this approach and nice 
discussion of other relevant aspects.

To illustrate this, consider the scalar descendant $\CO _{des}=P^\mu
\CO _\mu$ of a primary vector operator.  Writing \vectiipt\ as
\eqn\vectiipta{\ev{\CO _\mu (x)^\dagger\CO _\nu (0)}= {1\over(2\pi)^2}\,{C_V\over
2d}\left({1\over 2(d-2)}g_{\mu \nu}\partial ^2 -{1\over d-1}\partial
_\mu \partial _\nu\right) {1\over (x^2)^{(d-1)}},} 
we then have
\eqn\vectdoopt{\eqalign{\ev{\CO _{des} (x)^\dagger\CO _{des}
(0)}&= {1\over(2\pi)^2}\,C_V {(d-3)\over 4d(d-1)(d-2)}(\partial ^2)^2{1\over
(x^2)^{(d-1)}}\cr &=\, {1\over(2\pi)^2}\,{4C_V(d-1)(d-3)} {1\over (x^2)^{(d+1)}}.}}  
This is of the general form \scalariipt\ for the 2-point function of
the scalar operator, $\partial ^\mu O_\mu$, but has the wrong sign if
$1<d<3$. The range $d\le1$ can also be excluded; see Sec.~4.  Note
that it also follows from \vectdoopt\ that we can set $\partial ^\mu
\CO _\mu=0$ if, and only if the operator $\CO _\mu$ has dimension
$d=3$, exactly.  This fits with the fact that conserved currents have
$d=3$ exactly, with vanishing anomalous dimension.

\newsec{2-point functions in momentum space: ``unparticle propagators.''}

To consider the effect on SM scattering amplitudes of couplings
\smcft\ to a CFT, it is useful to Fourier transform the position space
CFT 2-point functions, {\it e.g.,} \scalariipt, \vectiipt, and
\tensoriipt, into momentum space propagators.  The Fourier integrals,
and their inverses, are generally singular, so continuations are
required.  First consider the scalar propagator. Evaluation of the
integral in Euclidean space yields
\eqn\fti{{1\over (2\pi )^2}{1\over (x^2)^d}={\Gamma (2-d)\over 4^{d-1}\Gamma (d)}\int {d^4k \over (2\pi )^4}e^{ik\cdot x} (k^2)^{d-2}.}  
Evaluation of the integral technically requires $ 0<{\rm Re}(d)<{5\over
4}$, but we assume it can be continued to all ${\rm Re}(d)>0$.

Similarly, for the position space  form \vectiipt\ for the two-point function of vector
operators of operator dimension $d$, we find the momentum space propagator
\eqn\vecft{{1\over (2\pi )^2}{g_{\mu\nu}-2x_\mu
x_\nu/x^2\over (x^2)^d} ={(d-1)\Gamma(2-d)\over
4^{d-1}\Gamma(d+1)}\int {d^4k\over (2\pi)^4} e^{ik\cdot x} (k^2)^{d-2}
\left[ g_{\mu\nu}-{2(d-2)\over d-1} {k_\mu k_\nu\over k^2}\right].}
Again, this computation was done in Euclidean space. To rotate to Minkowski
we take $k^4=-ik^0$ so $k^2\to -k^2$ (metric is $(+---)$). We need to
keep the contour, rotated clockwise from the imaginary axis toward the
real axis, off the poles or branch points on the real axis. These are
at $k^0=\pm|\vec k|$. So the contour goes under the real axis from
$-\infty$ to $-|\vec k|$ and above the real axis from $|\vec k|$ to
$\infty$. As usual this can be summarized by an $i\epsilon$
prescription, that the poles/branch points are at $k^0=\pm(|\vec
k|-i\epsilon)$, or $k^2+i\epsilon=0$. So we can write, for the
propagator of a dimension $d$ vector in Minkowski space (with
arbitrary normalization): 
\eqn\vecprop{\int d^4x e^{-ik\cdot x} \langle 0| T(O_\mu(x)O_\nu(0))|0 \rangle=-iC(-k^2-i\epsilon)^{d-3} \left[
k^2g_{\mu\nu}-{2(d-2)\over d-1}k_\mu k_\nu\right].}  
This differs from the propagator of \refs{\GeorgiEK, \GeorgiSI}\ in
the relative size of the terms.  The analogous expressions to \vecft\
for tensor operators can be similarly written. For the traceless, symmetric tensor,
\eqn\tensprop{\eqalign{
&\frac1{(2\pi)^2}{{\left(I_{\mu \lambda}(x)I_{\nu \sigma}(x)+ \mu\leftrightarrow\nu\right)-{1\over 2}
g_{\mu \nu}g_{\lambda \sigma}
}\over{ (x^2)^d}}=\cr
&\frac{\Gamma(2-d)}{4^{d-1}\Gamma(d+2)}\int
\frac{d^4k}{(2\pi)^4}e^{ikx}(k^2)^{d-2}
\bigg[d(d-1)\left(g_{\mu\lambda}g_{\nu\sigma}+\mu\leftrightarrow\nu\right)\cr
&+\frac12\left[4-d(d+1)\right]g_{\mu\nu}g_{\lambda\sigma}
-2(d-1)(d-2)\left(g_{\mu\lambda}\frac{k_\nu k_\sigma}{k^2}
  +g_{\mu\sigma}\frac{k_\nu k_\lambda}{k^2}+\mu\leftrightarrow\nu\right)\cr
&\qquad 
+4(d-2)\left(g_{\mu\nu}\frac{k_\lambda k_\sigma}{k^2}+g_{\lambda\sigma}\frac{k_\mu k_\nu}{k^2}\right) 
+8(d-2)(d-3)\frac{k_\mu k_\nu  k_\lambda k_\sigma}{(k^2)^2}\bigg]~,
}}
and for the anti-symmetric tensor:
\eqn\atensprop{\eqalign{
&\frac1{(2\pi)^2}{{\left(I_{\mu \lambda}(x)I_{\nu \sigma}(x)- \mu\leftrightarrow\nu\right)
}\over{ (x^2)^d}}=\qquad\cr
&\qquad-\frac{\Gamma(3-d)}{4^{d-1}\Gamma(d+1)}\int \frac{d^4k}{(2\pi)^4}e^{ikx}(k^2)^{d-2}
\bigg[\left(g_{\mu\lambda}g_{\nu\sigma}-\mu\leftrightarrow\nu\right)\cr
& \qquad\qquad\qquad\qquad\qquad\qquad\left.
-2\left(g_{\mu\lambda}\frac{k_\nu
      k_\sigma}{k^2} +g_{\nu\sigma}\frac{k_\mu
   k_\lambda}{k^2}-\mu\leftrightarrow\nu\right)
\right]~.
}}

The propagators in \vecft, \tensprop\ and \atensprop\ have an 
apparent singularity when
$d$ is an integer. This is similar to that of the propagator of
\refs{\GeorgiEK, \GeorgiSI} (using $\Gamma (d)\Gamma (1-d)=\pi/\sin
(\pi d)$).\ To illustrate what happens for integer $d$, consider the
case of $d=3$, where $\CO _\mu$ is a conserved current.  Taking $d\to
3$ in \vecft, there is a $1/(d-3)$ pole, which gives a local term,
proportional to $\partial^2\delta^{(4)}(x)$. Disregarding the local
terms (there are additional local terms from the finite piece, which
we also disregard), we have 
\eqn\intd{{1\over
(2\pi)^2}{g_{\mu\nu}-2x_\mu x_\nu/x^2\over (x^2)^{3}}= {1\over
48}\int {d^4k\over (2\pi)^4} e^{ik\cdot x} \left(k^2 g_{\mu\nu}-k_\mu
k_\nu\right) \ln(k^2).}  
As expected this has a log and is
conserved. The scale of the log is irrelevant since it corresponds to
a local term.  If we write the scale in the log as $\mu$, the $\mu$
dependence is cancelled by a contact counterterm ${1\over 24}(\partial
^2g_{\mu \nu}-\partial _\mu \partial _\nu)\delta ^4 (x)$ to the \RHS\ of
\intd.

The Fourier transforms \fti--\vecft\ are only valid for $x\ne0$. We
have just seen that in the special case $d=3$ an additional, local
term is required to make the propagator independent of the scale
$\mu$. For non-integer $d$, the propagator is non-singular, and it is
less clear if  a local counter-term is still needed.   In  section 5
we will show that the correct propagator must include a contact counterterm
even for non-integer $d$.

\newsec{Unitarity conditions, re-obtained physically}

In this section, we sketch  Mack's derivation \Mack\ of the constraints
on irreducible representations of the conformal group, in particular
for induced representations on Minkowski space. We then re-obtain the
unitarity constraints on CFTs from requiring unitarity of scattering
amplitudes of particles coupled to the CFT.   While our results are not
new, and in fact follow closely in the mathematics of Mack's original derivation, they give a new 
physical insight into  the origin of the constraints.

Consider group transformations, $\phi\to U\phi$. If on this vector
space we can find a group invariant, positive definite inner product
$(\cdot,\cdot)$, then $U^\dagger U=1$ on that space:
$$
(\phi_2,\phi_1)=(U\phi_2,U\phi_1)=(\phi_2,U^\dagger U\phi_1).
$$
For the conformal group, with primary operators $\phi $, consider the inner product
\eqn\innerprod{
(\phi_2,\phi_1)=\int d^4x\,\phi_2^\dagger(x)\Delta(x)\phi_1(x),}
where $\Delta(x)$ is just the Wightman function
\eqn\wightmans{
\Delta(x)=C\frac{P(x)}{(x^2+i\epsilon x^0)^d}.}
$C$ is a non-zero constant, and $P(x)$ is the polynomial in $I_{\mu\nu}$ appropriate for the
Lorentz spin of the representation, as in \scalariipt, \vectiipt,  and
\tensoriipt.  It is useful to consider
the Fourier transformed version (where the $\epsilon$ in \wightmans, amounting to  $x_0\to x_0+i\epsilon$, leads to $\theta (k_0)$): 
\eqn\innerprodf{(\phi_2,\phi_1)=\int d^4k\, \tilde\phi^\dagger_2(-k)\tilde\Delta(k)\tilde\phi_1(k),}
where $\tilde\Delta$ is the positive energy discontinuity across the cut
of the momentum space propagator.   By construction, this inner product is group invariant on primary fields. The representation is unitarity if \innerprodf\ is moreover convergent and positive definite.  As shown in \Mack, these conditions lead to the unitarity bounds \dineq\ for $(j_1,j_2)$ primary operators.   

As we now discuss, these same conditions follow from the physical requirement of positivity of
total scattering cross sections.  Using the optical theorem, we relate the CFT unitarity requirements directly to the positivity of the imaginary
part of the forward scattering amplitude, ${\rm Im}\,A_{fwd}>0$.  This corresponds to the positivity of \innerprodf. 

Consider first a scalar CFT operator $\CO (x)$, coupled 
to an external source $\chi$ through
a term in the lagrangian 
$$\CL \supset g\chi{\cal O} +\rm{h.c.}$$
The source $\chi$ may, of course, create or annihilate any number of
non-interacting (Standard Model) particles. 
The tree-level amplitude for $\chi\to\chi$, using \fti\ rotated to Minkowski space,  is then
$${\cal A}= g^2C_S|\chi|^2\frac{\Gamma(2-d)}{4^{d-1}\Gamma(d)}(-k^2-i\epsilon)^{d-2},$$
where $|\chi|^2$ is from the  creation and annihilation of
external particles with total momentum $k^\mu$.   We are interested in the imaginary part in the forward region, $s=k^2>0$.  We have dropped the t-channel contribution, because  it has no imaginary part
to leading order in $1/M^2$ (since $t<0$). 
Using $\Gamma(1-x)\Gamma(x)\sin(\pi x)=\pi$, the imaginary part of
the forward scattering amplitude is thus
\eqn\imAscal{
{\rm Im}\,{\cal A}_{\rm fwd}= {C_S\pi g^2 (d-1)\over4^{d-1}\Gamma(d)^2}|\chi |^2 \theta(k^0)\theta(k^2) (k^2)^{d-2}~.
}
By the optical theorem this must be positive, which implies
$$C_S(d-1)\ge0.$$
This, together with the condition that\foot{This was explained in the
discussion below  Eq.~\cpos\ but can also be seen directly from
$\vev{ \bar{\cal O}^\dagger\bar{\cal O}}>0$ with  $\bar{\cal
O}\equiv\int_{\cal R}d^4\!x\, {\cal O}$, where ${\cal R}$ is a compact
domain.} 
$C_S>0$, shows that $d<1$ is excluded.

The limit $d\to1$ requires some additional care. Despite the $d-1$ factor, \imAscal\ does not
identically vanish. 
Using $(d-1)\theta(k^2)/(k^2)^{2-d}\to\delta(k^2)$ as $d\to 1$ in
Eq.~\imAscal\ we have
\eqn\imAscald{
{\rm Im}\,{\cal A}_{\rm fwd}= C_S\pi g^2|\chi|^2 \theta(k^0)\delta(k^2) \qquad{\rm as }~d\to1}
(which is properly non-negative for $C_S>0$).  So $d=1$ corresponds
precisely to the exchange of a single scalar particle, with $k^2=0$,
corresponding precisely to a free field.

Next consider a CFT  vector operator, $\CO ^\mu$ of dimension $d$, which we couple to 
an external $\chi_\mu$ via a term $\CL \supset g\chi_\mu \CO ^\mu
+\rm{h.c}$.  Using \vecprop\ to compute the tree-level amplitude, this leads to a contribution with:
\eqn\imAvec{{\rm Im}\,{\cal
  A}_{{\rm fwd}}=-{g^2\pi C_V (d-1)^2\over 4^{d-1}d\,\Gamma^2(d)}\left[\chi\cdot\chi^\dagger-{2(d-2)\over d-1}{|\chi\cdot
    k|^2\over k^2}\right]\theta(k^2)\theta(k^0)(k^2)^{d-2}.}
Going to the center of mass frame $\vec k=0$, \imAvec\ is positive if 
$$\frac{C_V}{d}\left[|\vec \chi|^2+{d-3\over d-1}|\chi_0|^2\right]\geq 0,$$
and since this condition must hold for arbitrary $\chi_\mu$,  we have
${C_V}/{d}\geq0 $, and $(d-3)/(d-1)\ge0$.  This 
excludes $1<d<3$ and $d<0$ since\foot{Again, this follows from
$\vev{\bar{\cal O}^\dagger\bar{\cal O}}>0$ with  $\bar{\cal
O}\equiv\int_{\cal R}d^4\!x\,{\cal O}_\mu$, for any one fixed index
$\mu$.} $C_V>0$.   Finally, let us exclude also $0\le d<1$.  
We require the amplitude exists not just for in/out
plane waves but also for arbitrary but nice in/out
wave-functions. This means that we should be able to interpret
$\chi_\mu(k)$ and $\chi_\nu^\dagger(-k)$ as independent functions and
we require that the integral over $k$ converges. The light-cone singularity
of the integrand in
\eqn\dbound{\int d^4k\, \phi_1(k)\phi_2(-k)\frac1{(k^2)^a}}
is integrable for $a<1$. In our case above the $(k^2)^{d-2}(k_\mu
k_\nu/k^2)$ term gives the condition $3-d<1$ or $d>2$.   
Combining the above conditions, we have shown that we must have
$d\ge3$ for vector operators $\CO ^\mu$.   Note that, unlike the scalar case, the 
vector amplitude \imAvec\ is smooth when the unitarity bound is saturated, $d\to 3$, and all $k^2>0$
contribute. 

Whenever the unitarity bound inequalities \dineq\ are saturated, the representation of the conformal
group is smaller -- some descendants are set to zero \Mack.  In the scalar case, when $d\to 1$, 
this is reflected in the conversion of $\theta (k^2)\to \delta (k^2)$ in \imAscald. Indeed, 
\imAscald\ vanishes if $|\chi |^2=k^2$, since $k^2\delta (k^2)=0$; this corresponds to the fact that a scalar with $d=1$ has a scalar second descendant, $\partial ^2\CO $, with vanishing norm, requiring setting $\partial ^2 \CO =0$.    On the
other hand, for the vector case \imAvec, the limit where the unitarity bound is saturated, $d\to 3$,
is smooth,  and still involves $\theta(k^2)$ (rather than $\delta (k^2)$).  When the unitarity bound
is saturated for vectors, $d\to 3$, the operator with zero norm is $\partial _\mu \CO ^\mu =0$, which  corresponds to the vanishing of \imAvec\ for $\vec \chi=0$ when  $d=3$.

The arguments above can be easily generalized for other tensors.
Primaries with $j_1j_2=0$ are similar to the scalar case: when $d\to
j_1+j_2+1$, the $\theta (k^2)$ in ${\rm Im}\,\CA _{\rm fwd}$ becomes
$\delta (k^2)$, and some descendants are set to zero corresponding to
$k^2\delta (k^2)=0$.  On the other hand, primaries with $j_1j_2\neq 0$
are similar to the vector case: when $d\to j_1+j_2+2$, the behavior of
${\rm Im}\,\CA _{\rm fwd}$ is smooth, involving still $\theta (k^2)$
rather than $\delta (k^2)$; some first descendants have zero norm, and
hence vanish, due to the tensor structure of the terms.

Consider the case of the anti-symmetric tensor, which is $(j_1,j_2)=(1,0)+(0,1)$ (self-dual and anti-self dual).  The propagator, Eq.~\atensprop, has no $d$-dependence between
the different tensor structures, but the overall coefficient of the
absorptive part is, up to a positive quantity, $(d-2)(d-1)/d$.  Convergence excludes $d<1$ (as for
vectors),  so the unitary condition becomes
$d\ge2$.  The limit $d\to 2$ requires care, because of the overall factor of $(d-2)$: of 
the terms in the tensor
structure in \atensprop, only those with inverse powers of $k^2$ survive, since they give a contribution 
as $d\to 2$ via $(d-2)\theta(k^2)/(k^2)^{3-d}\to\delta(k^2)$.  For $d\to 2$, the $\delta (k^2)$ leads
(since $k^2\delta (k^2)=0$) to vanishing norm first descendants, which must be set to zero: $\epsilon^{\mu\nu\lambda\sigma}\partial_\nu
\CO _{\lambda\sigma}=\partial^\nu \CO_{\nu\mu}=0$. So $d=2$ precisely
corresponds to the exchange of a free $U(1)$ field strength tensor, $\CO_{\mu \nu}=F_{\mu \nu}$.  
Other $j_1j_2=0$ primary operators are similar. 

For the symmetric traceless tensor, $(j_1, j_2)=(1,1)$,  there are two differences from the
$(j_1, j_2)=(\half, \half)$ vector case above. First, convergence for wavepackets of the last term
in Eq.~\tensprop\ requires now $d>3$. And second, the tensor structure
has different $d$ dependence. Coupling to a tensor source,
$\chi_{\mu\nu}$ the condition from the $\chi_{00}$ or $\delta_{ij}{\rm
Tr}(\chi)$ components is $(d-4)(d-3)\ge0$. Convergence and positivity
lead to the condition $d\ge4$. As in the case of the vector, the limit where the unitarity bound
is saturated,  $d\to4$ is smooth.  There is a vanishing
norm first descendant (conservation law) $\partial^\mu O_{\mu\nu}=0$, due to the tensor structure
of the terms.  Other $j_1j_2\neq 0$ primary operators are similar.

\newsec{Weakly coupled CFT examples} 

\subsec{Illustration of the unitarity inequalities}

We here briefly demonstrate that the unitarity inequalities \dineq\ are
satisfied in a weakly coupled  Banks-Zaks \BZ\ type CFT.  Start with an
$SU(N_c)$ gauge theory with $N_f$ flavors of Dirac fermions $Q_f$,
$f=1\dots N_f$, in the fundamental representation.  Take the limit of
large $N_c$ and large $N_f$, holding $\epsilon \equiv {11\over
2}-{N_f\over N_c}$ fixed, with $0<\epsilon$ so that the theory is
asymptotically free, but just barely so, $\epsilon \ll 1$, and work to leading order in $\epsilon$ and $1/N_c$.  The beta
function for the $SU(N_c)$ gauge coupling $g$ in this limit is $\beta
(\alpha)=-b_1\alpha ^2+b_2\alpha ^3$, with $\alpha = g^2/4\pi$ and
$b_1=N_c\epsilon /3\pi$ and $b_2\approx 25N_c^2/16\pi ^2$.  
There is a zero of
$\beta (\alpha)$ at parametrically small 't Hooft coupling, $\alpha _*N_c\approx N_c b_1/b_2=16\pi \epsilon /75\ll 1$, so we expect perturbation theory to
be reliable, and work to lowest non-trivial order, $\CO (\epsilon)$. 

Consider first the scalar gauge invariant operators
\eqn\scalarbz{\CO _S = \delta ^{fg}\bar Q_f Q_g.}  To $\CO
(\epsilon)$, including the 1-loop anomalous dimension evaluated at the
RG fixed point, these operators have $d_S=3-{4\pi \epsilon \over 25}$.
Note that the anomalous dimension is negative, so $d_S<3$.  For small
$\epsilon$, these scalar operators remain well above\foot{ Indeed, the
gap equations conjecture of Refs.~\refs{\CohenSQ-\AppelquistDQ} is that the theory is only conformal if $d_S\geq 2$, and
below that instead has chiral symmetry breaking.  To better test the unitarity bound for 
scalar operators, consider a simple modification of the above example: add a gauge singlet, scalar field $\phi$, so $\CO
=\phi$ is a gauge invariant operator near the unitarity bound $d_S\geq
1$.  To make $\phi$ interacting, include a term $\CL \supset
h\phi \bar Q_fQ_g$, and note that for $\alpha _*N_c=\CO (\epsilon)$, where $\beta _g(g_*, h_*)=0$, the Yukawa coupling also has a RG fixed
point, $\beta _h(g_*,h_*)=0$ at $h_*^2=\CO (\epsilon)$.  As
required by the unitarity bound, the Yukawa interaction indeed leads
to $\phi$ having $\gamma _S\equiv d_S-1=+\CO (\epsilon)>0$. } the unitarity bound
$d_S\geq 1$.

Now consider the vector gauge invariant operators \eqn\vectorbz{J^\mu
= C^{fg}\bar Q^f\gamma ^\mu Q^g, \qquad \CO ^\mu = D^{fg}\bar
Q_f\gamma ^\mu \gamma ^5 Q_g,} where $C^{fg}$ and $D^{fg}$ are
constants.  Classically the operators \vectorbz\ all have $d=3$, so the 
unitarity bound $d_V\geq 3$ requires that their anomalous dimensions
must be non-negative.  The operators
$J^\mu$, for all $C^{fg}$, and $\CO ^\mu$ for the case of traceless
$D^{fg}$ are conserved currents, so they all have $d_V=3$ exactly,
without any quantum modification.  On the other hand, the operator
$\CO _\mu$, with $D^{fg}=\delta ^{fg}$, is not conserved -- it is
anomalous, $\partial _\mu \CO ^\mu =-{\alpha _*N_f\over 2\pi}\Theta$,
where $\Theta \equiv \Tr F_{\mu \nu}\widetilde F^{\mu \nu}$.  One can
verify that the anomalous current indeed has $\gamma _A\equiv d_V-3>0$, compatible with
the unitarity bound.  Writing the two-point function of the divergence of the
anomalous current using the general form \vectdoopt, 
\eqn\vectdes{\ev{\partial _\mu \CO ^\mu
(x) \partial _\nu \CO ^\nu (0)}={4C_V\gamma _A(2+\gamma _A)
\over
(x^2)^{4+\gamma _A}}= \left({\alpha _*N_f\over 2\pi
}\right)^2\ev{\Theta (x)\Theta (0)},}
with $C_V>0$.  Writing $\ev{\Theta (x)\Theta (0)}\approx C_S/(x^2)^4$ (working to lowest order
in perturbation theory), one
finds $C_S>0$, and then \vectdes\ yields
$\gamma _A \approx (\alpha _*N_f/2\pi )^2C_S/8C_V$, so $\gamma _A>0$, as expected.  This evaluation
of $\gamma _A$ corresponds to the three-loop diagram formed from two
copies of the 1-loop anomaly triangle, joined by their gluons in the
internal loop.

\subsec{Contact terms} In this section, we consider consider coupling
a weakly coupled CFT \BZ\ to the Standard Model, via a ultra-heavy
messenger sector, say a massive vector exchange of ultra-heavy mass
$M_U$ that couples a SM current, $j_\mu$, to a CFT vector operator,
$\CO _\mu$. As a concrete example, suppose $j_\mu = \bar e \gamma ^\mu
e+\bar \mu \gamma ^\mu \mu $.  We do not assume that $\CO _\mu$ is
conserved, {\it e.g.,} $\CO _\mu$ could be the anomalous current of
the previous subsection.  Integrating out the messenger at the scale
$M_U$, this induces
\eqn\opqi{{\cal
L}_{\rm{eff}}\supset D_1(M_U){1\over M_U^2}{\cal Q}_1,\qquad {\cal
Q}_1\equiv j_\mu \CO ^\mu.}  
Integrating out the ultra-heavy will generally also induce the contact terms
\eqn\opqii{{\cal L}_{\rm{eff}}\supset D_2(M_U){1\over M_U^2}\CQ _2
-D_2(M_U){1\over M_U^4}j_\mu \partial ^2j^\mu +\dots , \qquad \CQ
_2\equiv j_\mu j^\mu,} 
where the  two terms have the same
coefficients because they come from expanding the ultra-heavy
propagator $(M_U^2+\partial ^2)^{-1}=1/M_U^2-\partial ^2
/M_U^4+\dots$ 

In \refs{\GeorgiEK, \GeorgiSI}\ and following works on unparticles,
generally only the terms \opqi\ are included, without accounting for
the induced contact terms \opqii.  Effective contact interactions
associated with \opqi\ were considered in \BanderND.  The point of
this section will be to illustrate that the explicit contact terms in
\opqii\ are also needed.  In particular, as we will discuss, there is
important mixing between the term \opqi\ and the second term in
\opqii\ in, {\it e.g.,} $e^+e^-\to \mu ^+\mu ^-$ scattering.  For
exclusive scattering, the
first term in \opqii\ is a more important, leading order effect, which
would in general overwhelm the apparent unparticle effects. Of course,
such contact terms do not contribute directly to the peculiar phase
space effects associated with unparticle production. But the cross
section for direct unparticle production is $\sim 1/M_U^4$ while 
interference effects are $\sim 1/M_U^2$.
Observation of unparticle effects would likely come after observation
of these leading effects, except perhaps in models with a more
complicated messenger sector in which $D_1$, but not $D_2$, is
suppressed.

The amplitude we are interested in is
\eqn\heff{\langle\mu\mu| {\cal H}_{\rm{eff}}|ee\rangle, 
\qquad {\cal H}_{\rm{eff}}= {1\over M_U^4}\sum_{i=1,2}C_i{\cal O}_i,}
where the dimension 8 operators are
\eqn\opsare{\eqalign{ {\cal O}_1(0)&\equiv\int d^4x\,T({\cal Q}_1(x){\cal Q}_1(0)),\cr
{\cal O}_2(0)&\equiv -j^\mu(0)\partial^2 j_\mu(0).}}
We have included ${\cal O}_2$ because exchange of the BZ particles
between two insertions of ${\cal O}_1$ requires a $j^\mu j_\mu$
counterterm, as seen from the discussion after \intd.   There is also a contribution from ${\cal Q}_2$ but we
postpone including this to the end of this analysis, since there are
no subtleties associated with it.

Neglecting SM interactions ($j_\mu$ acts as a background field) the
insertion of ${\cal O}_1$ is just the CFT vector propagator $\ev{\CO _\mu (k)\CO _\nu (-k)}$, given by \vecprop, contracted with $j^\mu (k)j^\nu (-k)$.  (Because we chose a conserved current for $j^\mu$, only the 
$g_{\mu\nu}(-k^2-i\epsilon)^{d-2}$ term of the propagator
contributes). Denoting by $\Gamma_i$ the amputated Green function with
an insertion of ${\cal O}_i$, we have renormalization group equations
$$\left[{\partial\over \partial t}+\beta(g){\partial\over \partial
g}\right]\Gamma_i=-\gamma_{ij}(g)\Gamma_j$$ where $t=\ln\mu$ and
$\gamma$ is the anomalous dimension matrix for the operators ${\cal
O}_{1,2}$. We are interested in $d-3=\gamma_{11}\ne0$, so
$\CO_\mu$ is not conserved. Since we
neglect SM interactions, $\gamma_{21}=\gamma_{22}=0$, but
$\gamma_{12}\ne0$ and starts at order $g^0$, {\it i.e.,} it is present
in the free field theory case.  Close to the IR fixed point, $g=g_*$,
the solution is
\eqn\rgeCFTsol{\eqalign{
\Gamma_2(s,\mu)&=s\hat\Gamma_2 \cr
\Gamma_1(s,\mu)&=-\frac{\gamma_{12}(g_*)}{\gamma_{11}(g_*)}\Gamma_2
+\left[s\hat\Gamma_1+\frac{\gamma_{12}(g_*)}{\gamma_{11}(g_*)}\Gamma_2\right]
\left(\frac{\sqrt{s}}{\mu}\right)^{\!\!\gamma_{11}(g_*)},
}}
where $\hat\Gamma_i$ are constants and $s=k^2$.  The constant $\hat\Gamma_1$ is
fixed by the normalization of the operator $\CO _\mu$. If we could set
$\Gamma_2=0$ then $\Gamma_1(s,\mu)$ would precisely correspond to the
propagator \vecft\ with $d-3=\gamma_{11}(g_*)/2$. But this is not
possible. The $\Gamma_2$ terms in $\Gamma_1(s,\mu)$
in \rgeCFTsol\ are crucial in computing physical quantities
consistently, and can not be set to zero.  (While $\hat\Gamma_2$ is arbitrary, the product
$\gamma_{12}(g_*)\hat\Gamma_2$ is independent of any normalization
convention for ${\cal O}_2$.)   The contact term $\CO _2$ must be included.

To see this we return to the computation of the amplitude in \heff.
We run $\CH _{eff}$ in \heff\ to low energies, determining the running of
the  coefficient's $C_i$ by insisting that the amplitude be
$\mu$-independent:
\eqn\rgei{\left[{\partial\over \partial t}+\beta(g){\partial\over \partial g}\right]C_i=\gamma_{ji}(g)C_j.}
We can run these equations from the far UV, down to the IR.  There are matching conditions at $M_U$, 
$$C_1(M_U)=D_1^2(M_U)\qquad C_2(M_U)=D_2(M_U).$$
The second of these is explained following \opqii.    We do the RG running down in two steps.
In step 1, we run from the far UV, where the BZ gauge coupling $g\approx 0$, 
 down to the  ``dimensional transmutation scale,'' $\Lambda_U$, where the BZ gauge coupling runs near the IR fixed point value, $g  \approx g_*$, where $\beta (g_*)=0$.  In step 2, we integrate \rgei\ assuming that $g\approx g_*$, so we take $\beta (g)\approx 0$ and $\gamma(g)\approx\gamma(g_*)$.  In
step 1, the Green functions and coefficient functions are nearly
constant, since $g\approx 0$.  The one exception is
$C_2$ which runs even at $g=0$. Setting
$\gamma_{12}(g)\approx\gamma_{12}^{(0)}=$ constant, $C_2$ runs down to
\eqn\Cisolszero{C_2(\Lambda_U)\approx C_2(M_U)+\gamma_{12}^{(0)}C_1\ln(\Lambda_U/M_U).}
In step 2 the solution to \rgei\ is,
\eqn\Cisols{C_1(\mu)=\left({\mu\over\Lambda_U}\right)^{\!\!\!\gamma_{11}(g_*)}\!\!\!\!C_1(\Lambda_U),\qquad
C_2(\mu)=C_2(\Lambda_U)+{\gamma_{12}(g_*)\over \gamma_{11}(g_*)}
\left[\left({\mu\over \Lambda_U}\right)^{\!\!\!\gamma_{11}(g_*)}\!\!\!-1\right]C_1(\Lambda_U).}

Upon combining \rgeCFTsol\ and \Cisols,  it is easy to see that the
amplitude is explicitly $\mu$ independent, as expected.  But
had we ignored the contact term in the propagator ($\Gamma_2$ in the
second equation in 
\rgeCFTsol) there would have been residual $\mu$ dependence. As mentioned following \intd, a contact term is needed for $d_V=3$.  We have now shown that it is needed also for $d_V\neq 3$.  The above results indeed recover the $d_V=3$ case discussed after \intd\ upon taking the limit $\gamma_{11}\to0$ (the logs in \Cisolszero\ and \Cisols\
combine to give simply a $\ln(k^2/M_U^2)$ in the propagator of Eq.~\intd).

As mentioned at the start of this section, the contribution from the
operator ${\cal Q}_2$ to the $ee\to\mu\mu$ amplitude is much
bigger than the contributions of ${\cal O}_i$ which we have been
discussing. If $E$ is the center of mass energy then the ratio of the
 the CFT exchange amplitudes to the contact one is roughly
$${\CA _{\rm unparticle}\over \CA _{\rm contact}}\sim \frac{D_2^2}{D_1}
\left(\frac{E}{M_U}\right)^{\!\!2}\left(\frac{E}{\Lambda_U}\right)^{\!\!2(d-3)}.$$
Since $E<\Lambda_U<M_U$  both $E$-dependent factors
tend to suppress this ratio. Even if $\Lambda_U$ were close to 
$M_U$, as the energy $E$ is raised towards $M_U$ the effect of the
$s$-channel resonance becomes apparent and continues to overwhelm the
CFT exchange effect. The strategy for discovery should begin by
observation of the contact interactions, which would fix the scale
$M_U$, followed by high precision measurements to detect the small
residual unparticle effects.   In models with scalar unparticle exchange the contact interaction may
be less important, since $d_S$ can be lower, $d_S\ge1$.

\newsec{Amplitudes for vector unparticles, corrected}

Consider the decay $t\to q+U$ where $q=u$ or $c$ and $U$ stands for a
vector ``unparticle.''  Following Ref.~\GeorgiEK\ we
take the effective interaction
$${\lambda\over \Lambda^{d-1}}\,\bar q\gamma_\mu(1-\gamma_5)t
\CO ^\mu+\rm{h.c.}$$
but we take $\CO ^\mu$ to be a unitary, primary operator (\GeorgiEK\ takes a descendant $\CO ^\mu$).
The phase space is as in
Ref.~\GeorgiEK\ (we neglect the mass of the final state quark) but the
amplitude has a factor
\eqn\aprop{-g_{\mu\nu}+a k_\mu k_\nu/k^2}
from the vector matrix element. Here $k$ is the unparticle momentum and $a=2(d-2)/(d-1)$.
 The interesting range is $d>3$, where $a\neq 1$ (since $\partial _\mu \CO ^\mu \neq 0$).  We'll
keep $a$ as a parameter, to compare
easily with the incorrect result from using $a=1$.

\bigskip
\centerline{\epsfxsize=0.70\hsize\epsfbox{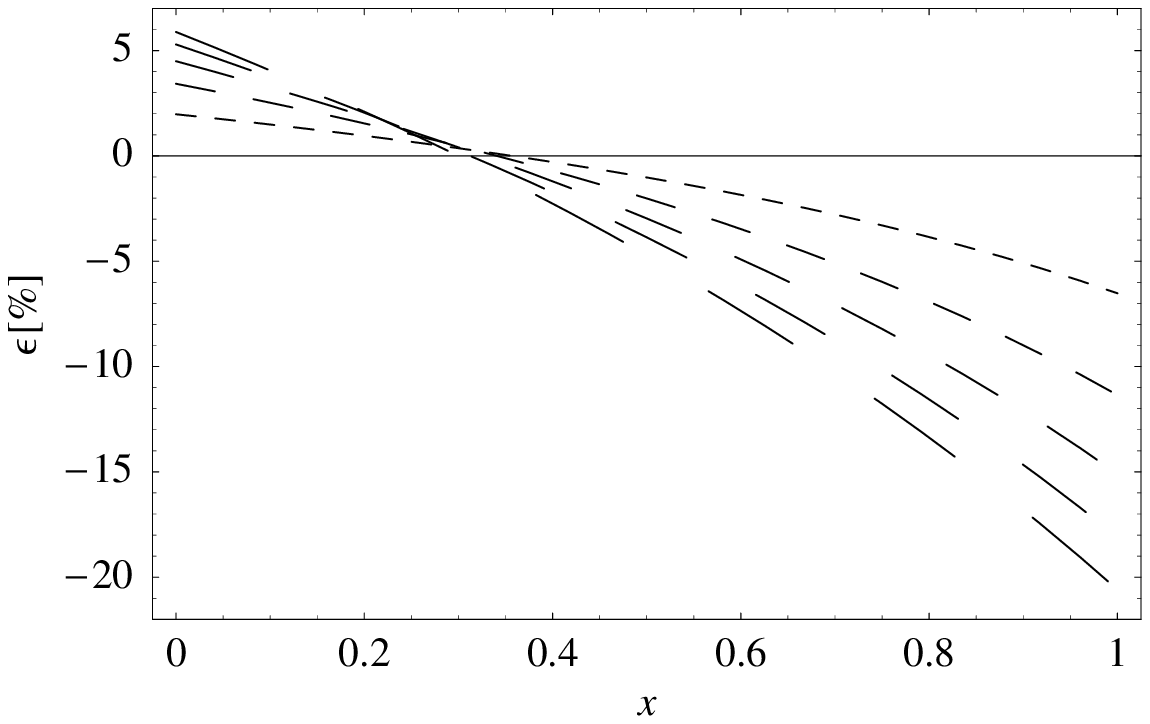}}
{\ninepoint\sl \baselineskip=2pt {\bf Figure 1:}
{\sl Fractional error $\epsilon$ in spectrum for the decay $t\to
   q+U$ as a function of $x=2E/m_t$ when the coefficient $a$ in
   \aprop\ is set to 1 instead of its correct value
   $2(d-2)/(d-1)$. The plots correspond to $d=3.2,3.4,3.6,3.8,4.0$
   with the length of dashing increasing with $d$.}}

Using the notation of Ref.~\GeorgiEK\ for the phase space and its
normalization factor we have
$${d\Gamma\over dx}=m_t{A_d\over 8\pi^2}\left({\lambda m_t^{d-1}\over\Lambda^{d-1}}\right)^2
x^2(1-x)^{d-2}\left[1-\half a+{(1-\half x)a\over 1-x}\right]~,$$
where $x=2E/m_t$. The spectrum is normalization independent, so it is
perhaps more interesting:
$$
{1\over \Gamma}{d\Gamma\over dx}={d(d^2-1)(d-2)\over 2d-4+a(d+1)}
x^2(1-x)^{d-2}\left[1-\half a+{(1-\half x)a\over 1-x}\right]~.$$
In Fig.1 we show the fractional error
$$
\epsilon\equiv\frac{\Delta\left(\frac1{\Gamma}\frac{d\Gamma}{dx}\right)}{\frac1{\Gamma}\frac{d\Gamma}{dx}}$$
where the difference is between the spectrum with the incorrect value,
$a=1$, and with the correct one, 
$a=2(d-2)/(d-1)$. 

\newsec{Summary} We have commented on several points concerning
CFTs\foot{Again, scale invariant theories are quite generally also
conformally invariant \PolchinskiDY.  There are no known examples
of 4d unitary theories which are scale but not conformally invariant,
and it is quite possible that such theories cannot exist.
Nevertheless, some of the unparticle literature misguidedly attempts
to ignore the constraints of conformal invariance, by restricting
their considerations to the hypothetical, perhaps non-existant, class
of theories which are scale invariant but not conformal.  For the sake
of completeness, we note that there are lower bounds on $d$ even
without using conformal symmetry.  The bound $d\geq 1$ for scalars,
reviewed in section 4, did not use the conformal symmetry.  More
generally, even without imposing conformal symmetry, unitarity and
convergence of \dbound\ for the term in the propagator with
maximum spin $j$ (maximum number of $k^2$'s in the denominator) gives
$d\geq j+1$.  And since $j_{max}=j_1+j_2$, this gives $d\geq
j_1+j_2+1$.  (Arbitrarily omitting the maximum $j$ terms from the
propagator would lead to a weaker inequality; {\it e.g.,} keeping only the
$j=0$ term, proportional to $g_{mu \nu}$'s, would weaken the bound to
$d\geq 1$.)  The stronger constraint of conformal symmetry only has
the effect of strengthening the bound for $j_1j_2\neq 0$, to $d\geq
j_1+j_2+2$.} which have been overlooked in the unparticle literature:
\lfm{1.} Unitarity imposes lower bounds on the dimensions of
operators (there is no upper bound, nor problematic behavior for 
integer dimensions). In particular $d\ge 3$ for vectors and $d\ge4$
for symmetric, traceless tensors. 
\lfm{2.} Only when the unitarity bound on the dimension is saturated
does the operator satisfy free field equations of motion or
conservation laws. Correspondingly, the tensor structure of the
propagators is modified.
\lfm{3.} Coupling the CFT to the SM via the exchange of an ultraheavy
particle necessarily introduces SM contact interactions, which generally
dominate over other unparticle interference effects.
Moreover,
CFT exchange induces additional SM contact interactions (which cure
the apparent problems with integer dimensions).
\bigskip
A number of interesting ideas and possible effects have been considered in the unparticle literature.  Where appropriate, the literature can be reanalyzed in light of the observations in this paper. 

\bigskip\bigskip

\noindent {\bf Acknowledgments}
We thank P. Puttayarat for discussions.  
The research of BG and KI is supported in part by Department of
  Energy under contract DOE-FG03-97ER40546 and that of IR by grants
  DOE-ER-40682-143 and DEAC02-6CH03000.

\listrefs
\end